\documentclass[twocolumn,resetfootnote]{aastex701}
\usepackage{CJK}
\newcommand{\ngc}{NGC\,4151}

\def\ixpe{IXPE}
\def\xmm{XMM-Newton}
\def\nustar{NuSTAR}

\def\spin{\hbox{$a^\ast$}}

\def\pdx{\hbox{$\Pi_{\rm X}$}}
\def\pax{\hbox{$\Psi_{\rm X}$}}

\newcommand{\rg}{\ensuremath{R_{\rm g}}}

\begin{document}

  \title{Disk reflection as the origin of the X-ray polarization of \ngc\ with IXPE}

\author[orcid=0000-0002-0273-218X]{E. Kammoun}
\affiliation{Cahill Center for Astronomy \& Astrophysics, California Institute of Technology, 1216 East California Boulevard, Pasadena, CA 91125, USA}
\email[show]{ekammoun@caltech.edu} 

\author[orcid=0000-0003-0079-1239]{M. Dov{\v c}iak} 
\affiliation{Astronomical Institute of the Czech Academy of Sciences, Bo{\v c}n{\'i} II 1401, CZ-14100 Prague, Czech Republic}
\email{}

\author[orcid=0000-0001-6264-140X]{J. Podgorn\'{y}} 
\affiliation{Astronomical Institute of the Czech Academy of Sciences, Bo{\v c}n{\'i} II 1401, CZ-14100 Prague, Czech Republic}
\email{}

\author[orcid=0000-0001-6264-140X]{I. E. Papadakis} 
\affiliation{Department of Physics and Institute of Theoretical and Computational Physics, University of Crete, 71003 Heraklion, Greece }
\affiliation{Institute of Astrophysics, FORTH, GR-71110 Heraklion, Greece}
\email{}
    
\author[orcid=0009-0005-7074-3699]{V. Binas-Valavanis} 
\affiliation{Department of Physics and Institute of Theoretical and Computational Physics, University of Crete, 71003 Heraklion, Greece }
\affiliation{Institute of Astrophysics, FORTH, GR-71110 Heraklion, Greece}
\email{}

\author[orcid=0000-0002-4622-4240]{S. Bianchi} 
\affiliation{Dipartimento di Matematica e Fisica, Università degli Studi Roma Tre, via della Vasca Navale 84, 00146 Roma, Italy}
\email{}

\author[orcid=0000-0002-9719-8740]{V. E. Gianolli} 
\affiliation{Department of Physics and Astronomy, Clemson University, Kinard Lab of Physics, Clemson, SC 29634, USA}
\affiliation{INAF-Osservatorio Astronomico di Brera, Via Brera 28, 20121 Milano, Italy}
\email{}

\author[orcid=0000-0001-9442-7897]{F. Ursini} 
\affiliation{Dipartimento di Matematica e Fisica, Università degli Studi Roma Tre, via della Vasca Navale 84, 00146 Roma, Italy}
\email{}

\author[0000-0003-3828-2448]{J. A. Garc\'{i}a}
\affiliation{X-ray Astrophysics Laboratory, NASA Goddard Space Flight Center, Greenbelt, MD 20771, USA}
\email{}

\begin{abstract}

    We present an X-ray spectro-polarimetric study of the nearby type-1 active galactic nucleus \ngc\ using two long \ixpe\ observations obtained in 2022 and 2024, supported by simultaneous XMM-Newton and NuSTAR spectroscopy. IXPE measures a polarization degree of $\sim 6-7\%$ above 4\,keV, with a polarization angle parallel to the radio jet, and a distinct low-energy component with a different angle, indicating at least two polarized components in the $2-8$\,keV band. Previous work interpreted the hard X-ray polarization as evidence for a radially extended slab-like corona. Here we test an alternative scenario in which the observed polarization is produced predominantly by relativistic reflection from an accretion disk illuminated by a compact, lamp-post-like corona. Using recently developed models, we fit the \ixpe\ Stokes spectra with a lamp-post plus distant-torus geometry, including partial-covering absorption and an additional soft polarized power-law component. We find that the data require a low coronal height ($h<9\,\rg$ at $3\sigma$) and a relatively large torus opening angle ($>45\degr$ at 3$\sigma$), while the disk reflection contributes $\sim 20\%$ of the $2-8$\,keV flux. The soft polarized component carries only $\sim 1-5\%$ of the flux but has a high polarization degree ($>10\%$) and a polarization angle around $20\degr$. The same configuration provides acceptable fits to the $0.4-79$\,keV \xmm\ and \nustar\ spectra, demonstrating that disk reprocessing by a compact corona can simultaneously account for both the polarization and broadband spectral properties of \ngc.

\end{abstract}


%

\section{Introduction}
\label{sec:intro}

The launch of the Imaging X-ray Polarimetry Explorer satellite \citep[\ixpe;][]{ixpe22} in 2021 opened a new window in the X-ray sky. For non-jetted and unobscured active galactic nuclei (AGN), \ixpe\ observations address a key question about the geometry of the X-ray corona, which remains unknown despite decades of X-ray observations. The X-ray corona is believed to be composed of hot relativistic electrons with a temperature ranging between a few tens and a few hundreds of keV \citep[see e.g.,][]{Fabian2015, Fabian2017, Tortosa2018, Akylas2021}. X-rays are then produced by Compton up-scattering the UV/optical thermal photons emitted by the accretion disk \citep[e.g.,][]{1979ApJ...229..318G, Sunyaev1980}. From X-ray spectroscopy only, it is hard to distinguish different coronal geometries. However, the expected polarization signal for the coronal emission is very sensitive to the geometry of the scattering material \citep[e.g.,][]{Schnittman2010, Tamborra18, Zhang2019, Ursini2022, Tagliacozzo2025}.

Since its launch, \ixpe\ observed a handful of bright nearby, non-jetted AGN. In obscured sources like Circinus galaxy and NGC\,1068, large polarization fractions with polarization angle perpendicular to the projected radio jet direction have been detected \citep[e.g.,][]{Ursini2023, Marin2024}. Such polarization state may be attributed to reprocessing inside a cold circumnuclear matter, which is extended from the equatorial plane and obscures the central regions. In unobscured sources, where the polarization is thought to trace, to a large extent, the structure of the innermost region, a statistically significant polarization detection has been reported in \ngc\ \citep{Gianolli23, Gianolli2024}, only. \cite{Ingram2023} found a similar result for IC\,4329A, although at a lower significance of $2.97\sigma$.
In the other sources observed with IXPE (MGC--05-23-16 and NGC\,2110), only upper limits have been reported \citep{Marinucci2022, Tagliacozzo2023, Chakraborty2025, Pal2025}.

\ngc\ was observed by \ixpe\ in 2022 and 2024 \citep{Gianolli23, Gianolli2024}. The polarization properties are consistent between the two observations. \cite{Gianolli2024} reported a polarization fraction of $4.5 \pm 0.9 \%$ with a polarization angle of $ 81 \degr \pm 6\degr$ east of north, combining both observations. The polarization angle is parallel to the extended radio emission in the source which is observed at a position angle (PA) of $\sim 77-83\degr$ \citep[e.g.,][]{Harrison1986, Ulvestad1998, Mundell2003, Williams2017}. Under the assumption of scattering by electrons within the corona as the polarizing mechanism, this suggests that the source of the polarized X-rays is located in a plane orthogonal to the jet direction. In this case, if all the polarization is attributed to the X-ray corona only, then the \ixpe\ results would favor a radially extended, slab-like, geometry over a more compact spherical corona. \cite{Gianolli2024} showed that two independent polarized components are acting below and above $\sim 4\,\rm keV$. The origin of the soft component was not identified. The hard polarization component was attributed to the X-ray corona. 

\xmm\ and \nustar\ observations were performed contemporaneously with the \ixpe\ observations, which enables a detailed spectro-polarimetric study of the source. \cite{Gianolli2024}  presented a phenomenological spectro-polarimetric analysis of these observations. In their model, they employed absorbed continuum plus neutral reflection models with separate polarization constants assigned to each spectral component. In their best-fit models that account for the observed polarization angle switch below 4\,keV (Models~2 and 3 in their Table~3), the polarization fraction attributed to the X-ray corona reaches $\sim12-19$\%, depending on the assumed polarization properties of the reflection component, which they fixed \textit{a priori} to either zero or 20\%. Such high values of the corona polarization can be challenging even for the slab geometry which predicts a maximum of $\sim 12\%$ depending on the inclination \citep{Ursini2022}. When a simplified model is adopted instead, consisting of an independently polarized power law and blackbody component without attempting to capture the full spectral complexity of the source, a lower coronal polarization fraction of $\sim$7\% is recovered (their Model~4), consistent with a slab-like geometry.

The studies so far have assumed polarization from the X-ray corona only. In this case, the measured values would exceed the expectations from a compact spherical (i.e., lamp-post) corona \citep[$\sim 2-3\%$;][]{Ursini2022}. However, if reflection off an accretion disk illuminated by a lamp-post corona is taken into consideration, larger polarization fractions would be expected. In this case, the polarization fraction will mainly depend on the corona height, disk properties, and the inclination of the system \citep[see e.g.,][]{Matt1993b,Podgorny23kyntokes}. This model can also naturally explain the polarization angle as the disk is thought to be roughly orthogonal to the radio jet. \cite{Gianolli2024} discussed the implication of the observed polarization to be produced by X-ray reflection within a rather phenomenological framework. In this work, we re-examine the observations of \ngc\ by testing disk reflection as a plausible explanation of the spectral and polarization properties of the source, using newly developed models which compute polarization in a self-consistent manner, taking into account all relativistic effects \citep{Podgorny23kyntokes}. 

The paper is organized as follows. In Section\,\ref{sec:obs}, we present the observations and data reduction. In Section\,\ref{sec:models}, we present the models used in this work. In Section\,\ref{sec:ixpe}, we present the analysis of the \ixpe\ spectro-polarimetric data. The spectral analysis of the \ixpe, \xmm, and \nustar\ observations is presented in Section\,\ref{sec:spectra}. Finally, we discuss our results and present our conclusions in Section\,\ref{sec:conclusion}.

\section{Observations and data reduction}
\label{sec:obs}

\begin{figure*}
\centering
\includegraphics[width=1\linewidth]{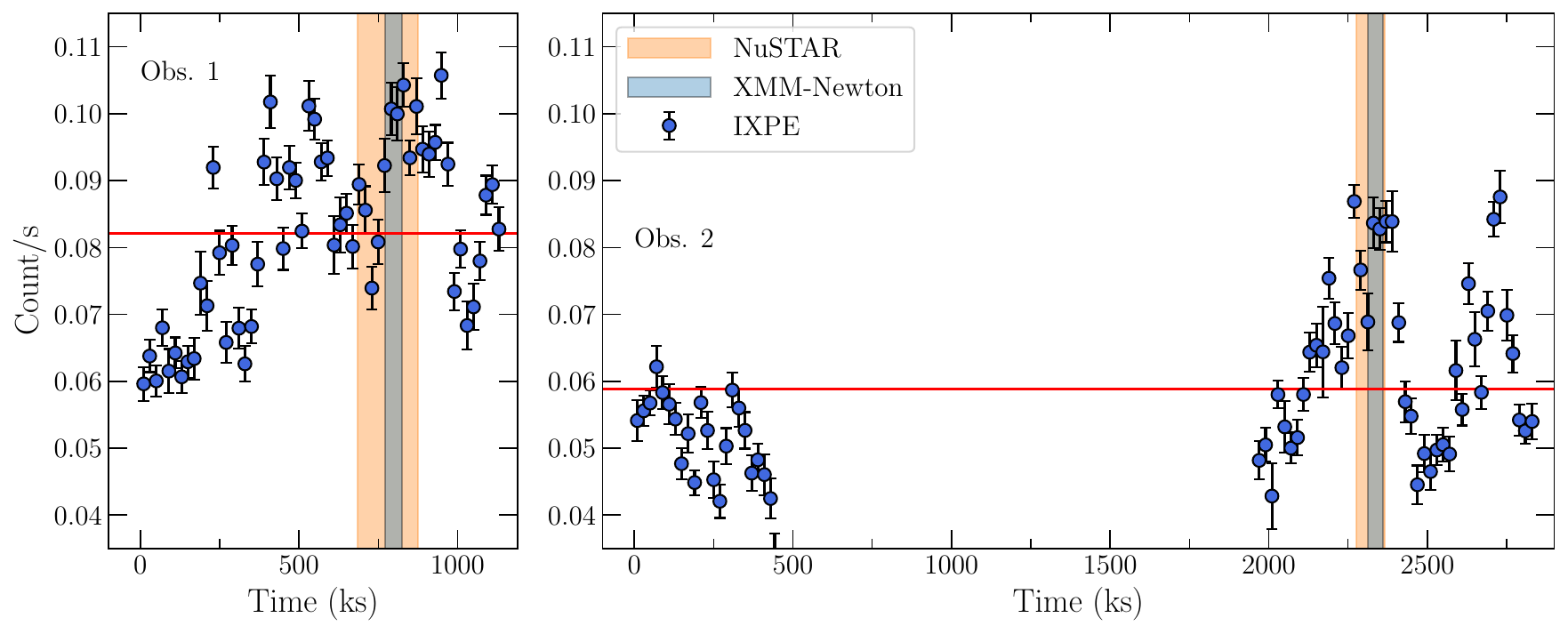}
\caption{The \ixpe\ $2-8$\,keV band light curve of \ngc\ for Obs\,1 and Obs\,2 (left and right panels, respectively), with a time bin of 20\,ks. For clarity, we show the light curves for DU1 only. The light curves of DU2 and DU3 are consistent with the ones shown in this figure. The horizontal lines indicate the average count rate in each of the observations. The shaded blue and orange regions show the periods when the XMM-Newton and NuSTAR observations took place, respectively.}
\label{fig:lcfig}
\end{figure*}


\subsection{IXPE}
\ixpe\ observed \ngc\ on December 8--21 2022 (ObsID\,02003101) for a net exposure of $\sim 632$ ksec and on April 23 to May 26 2024 (ObsID\,03006899) for a net exposure of $\sim 725$ ksec. We downloaded the data from the High Energy Astrophysics Science Archive Research Center (HEASARC)\footnote{See \url{https://heasarc.gsfc.nasa.gov/docs/ixpe/archive/}}. We have used Level 2 data sets with the source region defined in {\tt SAOimageDS9} v8.3 as a circle around the source centroid with an $80\arcsec$ radius. The background \citep{Dimarco2023} was extracted from an annulus centered on the source, with inner and outer radii of $120\arcsec$ and $210\arcsec$, respectively. The extraction of all three ($I,Q,U$) Stokes parameters in the original number of 150 energy channels for both the source and the background was performed using the {\tt xselect} tool from {\tt heasoft} v6.33.2 with simple weighting \citep[i.e., {\tt extract "SPECT" stokes=SIMPLE} was used;][]{Dimarco2022}. The IXPE responses\footnote{\url{https://heasarc.gsfc.nasa.gov/docs/ixpe/caldb/}} v13 from 28 February 2024 were used . The IXPE {\tt arf} and {\tt mrf} response files were computed with the {\tt ixpecalcarf} tool. 

Fig.\,\ref{fig:lcfig} shows the $2-8$\,keV band light curve of \ngc\ during the two \ixpe\ observations. The source shows clear inter- and intra-observation variability. During Obs\,1 the source flux was higher than Obs\,2 by $\sim 30\%$ on average. The flux varies by a  max/min factor of\,$\sim 1.7$ in Obs\,1, and $\sim 2.2$ in Obs\,2.

\subsection{XMM-Newton}
\xmm\ observed \ngc\ on December 17 2022 (ObsID 0921160201) and on May 19--20 2024 (ObsID 0934990401). The \xmm\ European Photon Imaging Camera (EPIC) pn operated in the Small Window mode during both observations. We reprocessed the \xmm\ data with the scientific analysis system (SAS) v22.0.0, and calibrated them using the latest calibration files as of February 2025. The observation data files (ODF) were downloaded from the \xmm\ Science Archive (XSA), and the {\tt epproc} command was used to create event files. We selected single and double pixels for the event patterns ($0-4$), as well as ``flag"=0 (i.e., the source region contains no bad pixels). Light curves and spectra were extracted using a circular source region with a radius of $40\arcsec$. We selected two rectangular background regions in places where the source photon counts were negligible for each observation. The area of the two background regions was 2.8 and 1.9 times larger than the area of the source region for the first and second observations, respectively.

In order to check for high–energy flaring particle background, we created light curves in the $10-12\,$keV energy band (using event pattern 0, and flag=0). Both observations of \ngc\ had flares present in small periods in the $10-12\,$keV band. However, these flares do not appear in the background subtracted light curves in the $0.3-10\,$keV, as well as in the $0.3-2$ and $2-10$\,keV bands. Therefore, we also considered data from these short periods. For both observations the pile-up effects were found negligible in the $0.3-10\,$keV band, using {\tt epatplot}. We extracted spectra in the $0.3-10\,$keV band, and created redistribution matrix files (rmf) and auxiliary response files (arf) using {\tt rmfgen} and {\tt arfgen}, respectively. Corrections to the effective area removing residuals between simultaneous fits of \xmm\ and \nustar\ observations were applied with {\tt applyabsfluxcorr}. Lastly, we binned all the spectra with the {\tt ftgrouppha} from NASA’s HEASARC software package {\tt ftools}using ``{\tt grouptype=optmin}'' and ``{\tt groupscale=20}''. These options set up the binning using the optimal binning scheme of \cite{Kaastra2016} with the additional requirement of a minimum number of 20 counts in each bin.

\subsection{NuSTAR}
\nustar\ \citep{Harrison2013} observed \ngc\ with its two coaligned X-ray telescopes, focal plane modules A (FPMA) and B (FPMB), on December 16--18 2022 (ObsID 60901003002) and on May 19--20 2024 (ObsID 60902010004). Data were retreived from the \nustar\ Archive\footnote{\href{https://heasarc.gsfc.nasa.gov/docs/nustar/nustar_archive.html}{https://heasarc.gsfc.nasa.gov/docs/nustar/nustar\_archive.html}}, and analyzed using CALDBv20250203. Analysis was performed with the \nustar\ Data Analysis Software (NuSTARDAS) v2.1.5. Level 2 calibrated and clean event files were produced with the {\tt nupipeline} script. Since the elapsed time of the \nustar\ observations was larger than that of \xmm, we created good-time interval files (GTIs) using {\tt xselect} ({\tt heasoft} v6.35.1), that include only the simultaneous parts. The shaded regions in Fig.\,\ref{fig:lcfig} show the periods in which the XMM-Newton and NuSTAR observations took place.

We extracted the $3-80\,$keV spectra, arf, and rmf from the clean event files with the {\tt nuproducts} module for both \nustar\ FPMA and FPMB. The source extraction region was circular with radius equal to $120\arcsec$. One rectangular background region was selected in areas where the photon counts of the source were negligible for each observation. The area of the background region was 3.4 and 2.6 times larger than the area of the source region for the first and second observations, respectively. Finally, the spectra were binned using the same binning method as the one used for the \xmm\ spectra.

We note that three \xmm\ and \nustar\ observations were performed during the \ixpe\ observation of 2024. Given that we require strict simultaneity between \xmm\ and \nustar\ for our analysis, we present here the results from the observation with the longest \xmm\ exposure ($\sim 35$\,ks net; as opposed to $\sim 9-20$\,ks net exposure for the other two observations). It is also worth noting that this observation was not studied by \cite{Gianolli2024} who analyzed data from the other two observations in 2024. Adding these two observations does not change any of our conclusions. The 2022 observations (Obs\,1) are the same as the ones presented by \cite{Gianolli23}.


\section{Spectro-polarization models}
\label{sec:models}

In this work, we test the scenario of a compact spherical X-ray corona located on the rotation axis of the central black hole irradiating the accretion disk (commonly known as the lamp-post geometry). In this scenario, we assume that all of the observed polarization is caused by reprocessing the X-ray radiation by an accretion disk and/or a more distant material (e.g., the torus). The corona emission and the related disk reflection is modeled using {\tt KYNSTOKES}\footnote{\url{https://projects.asu.cas.cz/dovciak/kynstokes}} \citep{Dovciak2011,Podgorny23kyntokes}. The distant reprocessing is modeled using {\tt stokes\_torus}\footnote{\url{https://github.com/jpodgorny/stokes_torus}} \citep{Podgorny2024}. Both of these models are available for data fitting in {\tt XSPEC} \citep{Arnaud1996}.

\begin{figure}
\centering
\includegraphics[width=\linewidth]{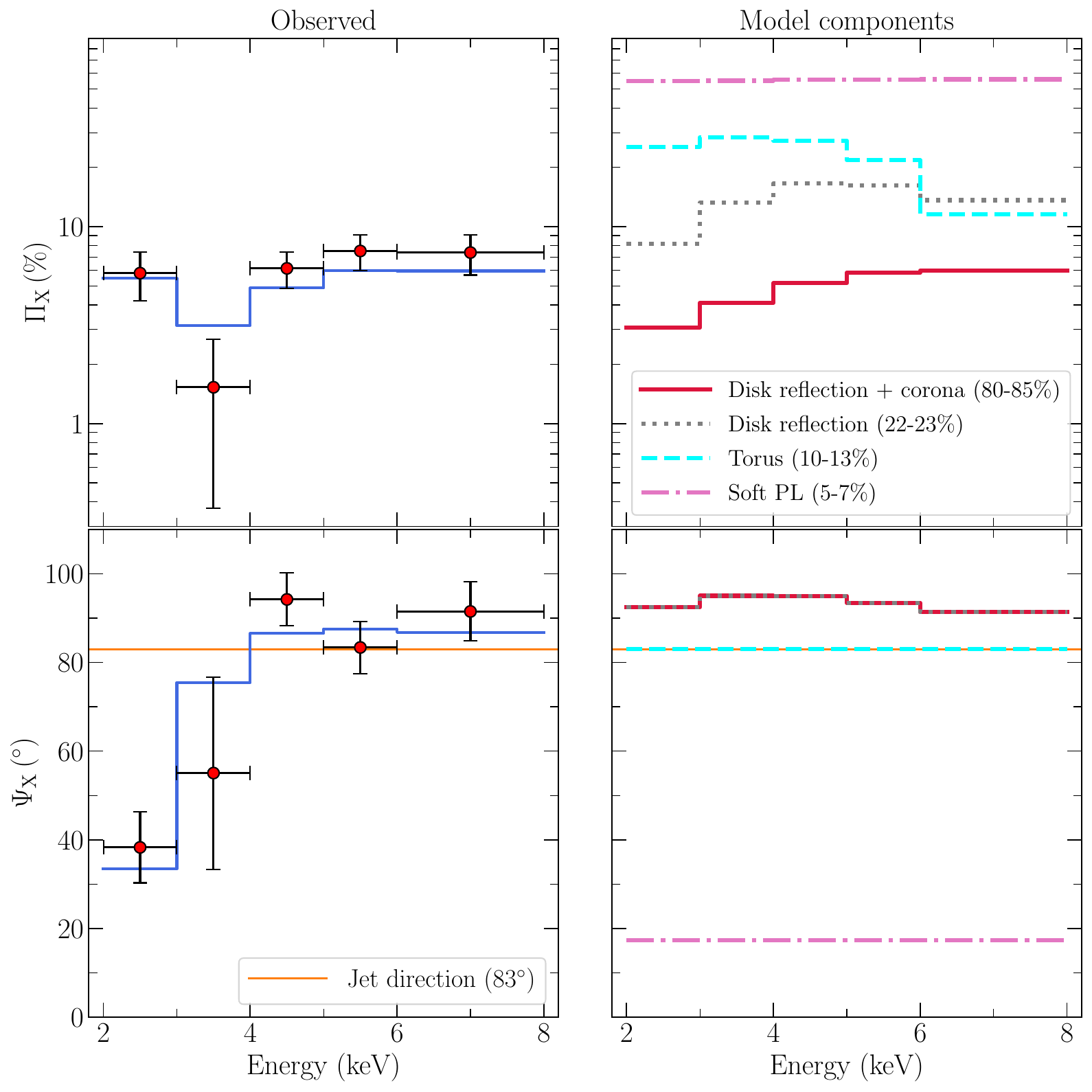}

\caption{Polarization degree and angle versus energy in \ngc\ (top and bottom panels, respectively). The left column shows the observed polarization parameters combining the two observations together with the best-fit model (blue lines), and assuming $\rm PA=83\degr$ (horizontal dashed line in the bottom panels) and a spin parameter $a^\ast = 0.75$. The right column shows the individual best-fit model components (see text for details). The percentages listed in parentheses show the contribution of each component to the total observed flux.}
\label{fig:polvsenergy}
\end{figure}

\subsection{{\tt KYNSTOKES}}

{\tt KYNSTOKES} calculates the spectro-polarimetric properties  of the primary and reflected X-ray emission from the inner region of an accretion disk, as will be seen by a distant observer. We present below a brief summary of how the model works. The model was first introduced in \cite{Dovciak2011} and was extended for polarization of the primary source and a detailed treatment of disk ionisation in \cite{Podgorny23kyntokes}. {\tt KYNSTOKES} can also  operate, to a limited extent, in the slab corona geometry. In this work, we consider only the lamp-post geometry. We therefore assume that the X-ray corona is a point-like source located on the rotational axis of the accretion disk at some height, $h$, above the BH, and it emits isotropically (in its rest frame), a power-law-like energy spectrum. For our purposes, we assume the primary emission to be unpolarized in accordance with \cite{Ursini2022}. In this case, X-rays detected by a distant observer are polarized at an angle nearly parallel to the axis of symmetry of the system via reflection of unpolarized photons off the accretion disk \citep{Podgorny23kyntokes}.

The disk is geometrically thin and optically thick, it has a Keplerian velocity profile and a constant radial and vertical density profile, with $n_{\rm e} = 10^{15}\,\rm cm^{-3}$. It is also partially ionized depending on the local X-ray incident flux. We assume that the disk extends down to the innermost stable circular orbit (ISCO). 

The model uses the local reflection tables of \cite{Podgorny22} that were computed by the radiative transfer code {\tt TITAN} \citep{Dumont03} to obtain the vertical ionization structure of a disk atmosphere and the 3D Monte Carlo code {\tt STOKES} \citep{Goosmann2007, Marin12, Marin15, Marin2018} which includes the physics of reprocessing (absorption, line re-emission, and scattering) to produce a complete spectro-polarimetric output. The tables include numerically simulated polarization properties of the disk X--ray reflection component (the continuum and major spectral lines) in the rest frame with the accretion disk. {\tt KYNSTOKES} then performs the necessary interpolation of the local tables and integrates them over the geometrically thin accretion disk in the equatorial plane using all special and general-relativistic effects apart from returning radiation effects (i.e.m secondary reflections due to light bending). It computes the light trajectories from the lamp to the disk, from the lamp to the observer, as well as from the disk to the observer. We note that the model does not account for possible polarization effects from the outer components in AGN, such as jets, the broad and narrow-line regions, polar winds, or the putative dusty torus. 

The energy-dependent Stokes parameters for a distant observer are then provided by {\tt KYNSTOKES} for a range of parameters, such as the mass and the spin of the BH, the disk inclination, the height of the primary source above the BH, the $2-10$\,keV luminosity of the primary source (normalized to the Eddington luminosity, $L_{2-10}/L_{\rm Edd}$), the primary power-law photon index, the primary polarization state, and the inner and outer radii of the disk.

\subsection{{\tt stokes\_torus}}

{\tt stokes\_torus} (v1.1) computes the spectro-polarimeteric properties obtained by reprocessing an X-ray power-law emission (of arbitrary incident polarization) by a nearly neutral opaque toroidal structure, as presented in \cite{Podgorny2024}. The model represents an optically thick AGN torus illuminated by a central compact hot X-ray corona. All of the model components are assumed to be static. The central source is assumed to be an isotropic point-like emission. The energy-dependent Stokes parameters of the reflected component for a distant observer are provided by {\tt stokes\_torus} for a range of parameters, such as the incident photon index, and the torus inclination and opening angle. It also takes into account the polarization degree and angle of the incident radiation. Assuming a reasonably-polarized (a few percent) incident radiation, which could be caused by the accretion disk reflected emission, does not affect our results. Thus, we assume that the incident radiation is unpolarized for {\tt stokes\_torus}.

\section{Spectro-polarimetric results}
\label{sec:ixpe}

To increase the polarization signal-to-noise and to prevent the fits to be more driven by the total flux (Stokes parameter $I$), we binned the Stokes parameters ($I, Q, U$) in five energy bins (in the $2-8$\,keV energy band), for each of the three detectors using the FTOOLS \texttt{grppha}. Since the Stokes parameters are sums over individual photon events, binning to an adequate number of counts per bin renders their distributions approximately Gaussian \citep{Kislat2015}, so that $\chi^2$ minimization is statistically valid. We do not combine the two observations but we model them simultaneously. This results in a total of 90 data points (45 points per observation). We find that the polarization properties are consistent between the two observations. For visualization purposes only, we show in Fig.\,\ref{fig:polvsenergy} the polarization fraction ($\Pi_{\rm X}$) and polarization angle ($\Psi_{\rm X}$) by combining all of the detectors and both observations, using the {\tt XSPEC} command ``{\tt setplot group}". 

Fig.\,\ref{fig:polvsenergy} shows that \pax\ changes with energy starting at a low value below 4\,keV and reaching a high and constant value above 4\,keV that is consistent with the jet orientation of $83\degr$ (shown as a horizontal line in the bottom panel of Fig.\,\ref{fig:polvsenergy}). As for \pdx, our results show a high degree of polarization (more than 4\%) in all energy bins except the $3-4$\,keV bin where \pdx\ drops to $1.53\pm1.15 \%$. These findings are consistent with the results by \cite{Gianolli2024}.

We fit the ($I,Q,U$) Stokes parameters as a function of energy for each of the detectors and for both observations simultaneously, leaving a calibration constant free to vary between the three detectors. We assume the following model in {\tt XSPEC} parlance:

$${\tt model = zpcfabs \times zpcfabs \times KYNSTOKES} $$
$${\tt + stokes\_torus + constant \times polconst \times powerlaw}.$$

This model consists of two partially covering neutral absorbers ({\tt zpcfabs}) that act on the corona \citep{Gianolli23,Gianolli2024} and disk emission ({\tt KYNSTOKES}). In addition, we account for the reflection from nearly neutral material using {\tt stokes\_torus}. We added the term (${\tt constant \times polconst \times powerlaw}$) to account for an additional component that is required to fit the polarization signal at energies below $\sim 4\,\rm keV$. The need for such a component, of unknown origin, has also been reported by \cite{Gianolli2024}. We parametrize the flux of this component as a fraction of the corona power law flux ($C_{\rm PL}$), which has the same photon index. The free parameters are the black hole spin ($a^\ast$), the corona height ($h$), the photon index ($\Gamma$), the X-ray $2-10$\,keV luminosity in units of Eddington luminosity ($L_{\rm 2-10}/L_{\rm Edd}$), the torus opening angle ($\theta_{\rm torus}$), the normalization of {\tt stokes\_torus} ($\rm Norm_{torus}$), the fraction of flux that goes into the additional power-law component ($C_{\rm PL}$), and the polarization fraction and angle of this component ($\Pi_{\rm PL}$ and $\Psi_{\rm PL}$, respectively). We left the values of $\Gamma$ and $L_{\rm 2-10}/L_{\rm Edd}$ free to vary between the two observations. 

The number of free parameters is large and we cannot constrain all of them by fitting the data from \ixpe, \xmm, and \nustar. For this reason, we tried to keep some of them fixed to reasonable values during the fitting process, as we explain below. First, we fix the BH mass to $M_{\rm BH} = 1.7\times 10^7\,\rm M_\odot$ \citep{Bentz2022} and the Eddington ratio to $\dot{m}/\dot{m}_{\rm Edd} = 0.02$ \citep{Lubinski2016}. We also fixed the Fe abundance to the solar value in all reflection models. 

\begin{figure}
\centering
\includegraphics[width=0.99\linewidth]{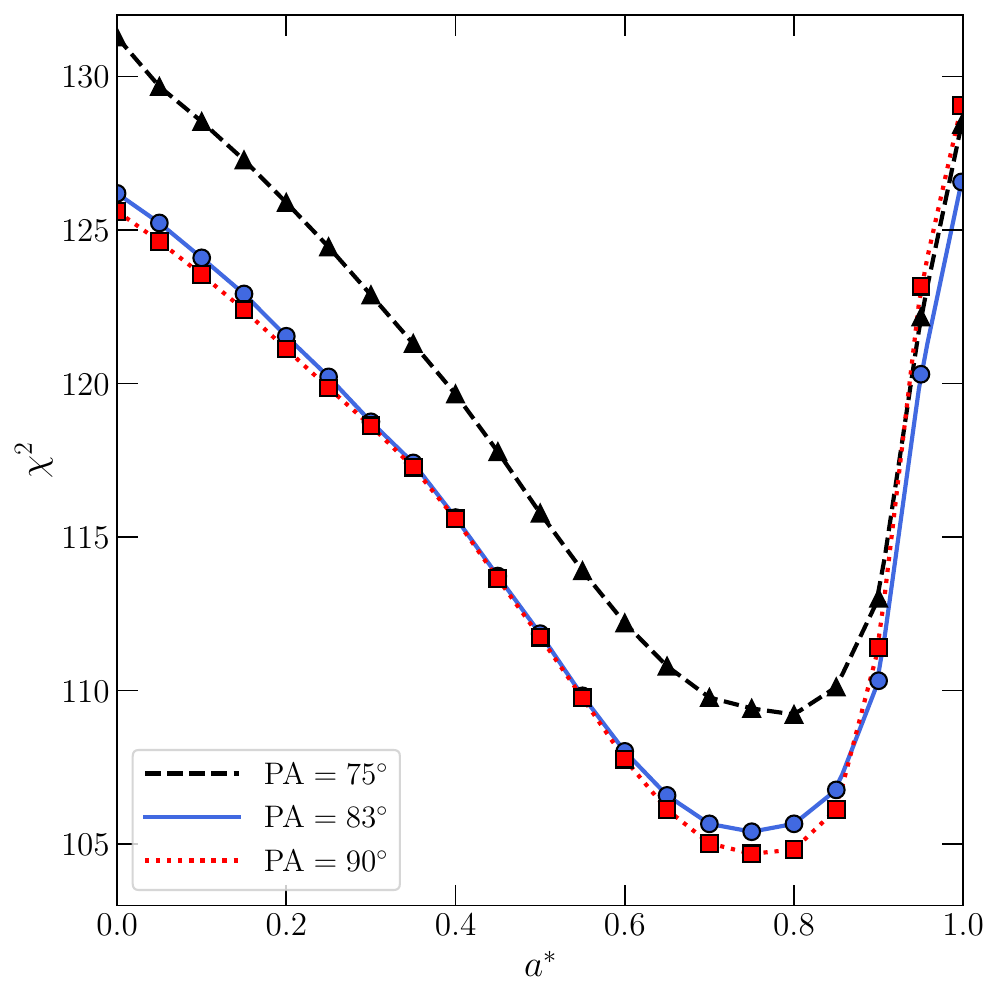}

\caption{Distribution of $\chi^2$ as a function of spin obtained by fitting the Stokes parameters assuming a PA of $75\degr, 83\degr,$ and $90\degr$ (black, blue, red, respectively). }
\label{fig:chi2_spin}
\end{figure}


Both \texttt{KYNSTOKES} and \texttt{stokes\_torus} compute the polarization angle with respect to the rotation axis of the BH (which is vertical to the accretion disk plane). However, the polarization angle that a distant observer will detect also depends on the orientation of the system, i.e., on the angle between the rotation axis with respect to the north direction (i.e., on the position angle, PA, of the system). Various PA values have been reported in the literature for \ngc\ based on the orientation of the radio jet in this source \citep[see e.g.,][]{Ulvestad1998, Mundell2003}. These values vary between $\sim 77\degr$ and $90\degr$. \cite{Gianolli23, Gianolli2024} adopted $\rm PA = 83\degr$. Thus, we first considered three values of PA as $\rm PA = 75\degr, 83\degr$, and 90$\degr$. For each of these values, we fitted the Stokes parameters as a function of energy by fixing the spin to values between 0 and 0.998 with a step of 0.05 (and letting all other parameters free to vary). The distribution of $\chi^2$ as a function of spin for each value of PA is shown in Fig.\,\ref{fig:chi2_spin}. This figure shows that the fit improves for a larger PA. The best-fits for PA of $83\degr$ and $90\degr$ are comparable, with the latter being slightly better. Thus, we adopt this value to fit the \ixpe\ data, similar to \cite{Gianolli2024}.

Furthermore, the $\chi^2$ plot in Fig.\,\ref{fig:chi2_spin} can also be used to infer the best-fit \spin\ and its uncertainty, which is $0.75^{+0.19}_{-0.40}$ (the uncertainty was calculated for $\Delta\chi^2=11.75$, which is appropriate for the case of 1$\sigma$ errors for 10 free parameters). The result indicates that a relatively wide range of spin values can fit the data well. Since the BH spin is not very constrained, we fixed it at 0.75 in all model fits, to increase the accuracy of the other model parameter values.

 We consider the inclination of the accretion disk and of the torus to be the same, and we keep it fixed during model fits. Considerable neutral absorption, which affects the soft X-ray emission from the central source, is almost always detected in \ngc. This implies that the inclination angle may be large in this system, so that we observe it through the upper layers of the putative torus. If the opening angle of the torus is of the order of 40\degr, then the inclination angle can be as high as 60\degr\ in \ngc. This seems to be supported by recent reports from the modeling of the broad-line region in this source, which imply an inclination angle of 60$\degr$ \citep{Bentz2022}. If the accretion disk and the BLR share the same axis of symmetry, then it would be reasonable to adopt this value for the model fitting. 

To further investigate this issue, we fitted the data for $\theta$ between 45 and 80 degrees with a step of $\Delta\theta=5\degr$ (we let all parameters free during the fits, except the mass and spin of the BH, as well as the accretion rate, as discussed above). The best-fit $\chi^2$ values decrease from 115 to 104 (for 77 degrees of freedom; dof). The respective null hypothesis probability  ($p_{\rm null}$) increases from 0.002 to 0.02, respectively. If we assume that the models with $p_{\rm null}>0.01$ are statistically acceptable, then the results suggest that the X-ray reflection in the accretion disk can account for the observed polarization properties of \ngc, but only for inclination angles larger than 55\degr. Since it may be rather non-physical, even for a Seyfert\,1.5 such as \ngc, to assume inclination angles larger than $60\degr-70\degr$, we decided to keep the inclination angle fixed at 60\degr during the model fits. We note that the inclination of the innermost accretion disk does not need to coincide with larger-scale estimates from the BLR/torus or the host galaxy if the inner and outer disk axes are misaligned (e.g., through a disk warp). Nonetheless, the analysis above shows that the high polarization degree measured by \ixpe\ independently drives the inner-disk inclination to $i \gtrsim 55\degr$, regardless of the larger-scale geometry, lending support to the adopted value.

Given the limited energy band, the covering fractions of the absorbers are both pegged at 1 and the column density of one of the absorber was consistent in both observations. Thus, we fixed the covering fractions and tied the column density of one of the absorbers. Given the poor constraints of the covering fractions, we could replace the two absorber with a single fully covering absorber at a larger column density ($N_{\rm H}$). Doing so does not affect the other parameters or the fit statistics. However, we decided to preserve a configuration with two absorbers that is more in line with the model used later to fit the energy spectra (Section\,\ref{sec:xmm}). The final model consists of 13 free parameters. 

We obtained a best fit with $\chi^2 = 105$ for 77 degrees of freedom ($p_{\rm null}=0.02$). The best-fit parameters are listed in Table \ref{tab:Polpar}. During this fit, we kept the corona height the same in both observations. In principle, we could assume that the height of the corona is not the same in both observations. They have different luminosities and different photon indices, so it would be reasonable to expect that there have been changes to the coronal properties, such as the corona height. We let the corona height to be different in the two observations, but the quality of the fit does not improve ($\Delta\chi^2=-2$ for one additional free parameter), and the best-fit corona heights are consistent within the errors (which are very large).

\begin{figure*}
\centering
\includegraphics[width=0.99\linewidth]{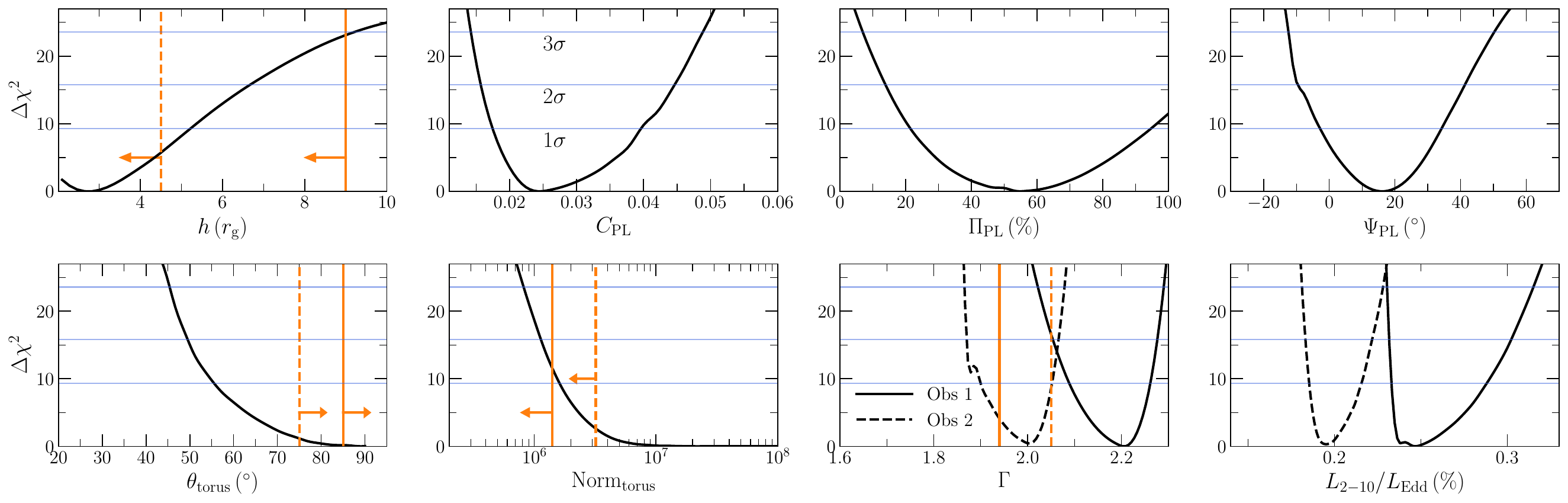}

\caption{Plots of $\Delta \chi^2$ as a function of each of the free parameters. Solid and dashed lines correspond to Obs\,1 and Obs\,2, respectively. Parameters with a single solid line were assumed to be constant between observations. The horizontal lines show the $1\sigma, 2\sigma,$ and $3\sigma$ confidence levels for 8 free parameters. The vertical orange lines correspond to the best-fit values obtained by modeling the \xmm\ and \nustar\ spectra.}
\label{fig:chi2_parameters} 
\end{figure*}


\begin{table*}[]
    \centering
    \begin{tabular}{llll|l}
    \hline \hline
        	&		&	\ixpe	&		& XMM+NuSTAR\\ 
        Parameter	&	Best-fit value	&	$1\sigma$ range	&	$3\sigma$ range	& 	$3\sigma$ range \\ \hline
        Black hole spin $a^\ast$  & $0.75^{\rm fixed}$    &   & & $[0,0.998]$\\
        Disk inclination $i$ ($\degr$)  & $60^{\rm fixed}$    &   & &  $60^{\rm fixed}$ \\
        Corona height $h\,(\rg)$	&	2.5	&	$< 5 $	&	$< 9 $	& Obs\,1: $<9$\\	
        	&		&		&		& Obs\,2: $<4.5$\\	
        Torus opening angle $\theta_{\rm torus}\,(\degr)$	&	89	&	$> 55$	&	$> 45$	& Obs\,1: $>85$ \\	
        	&		&		&		& Obs\,2: $>75$\\	
        Torus reflection normalization 	&	$10$	&	$>0.16$	&	$>0.08$	& Obs\,1: $<0.14$ \\	
        $(10^7\,\rm photons\,s^{-1}\, cm^{-2}\,keV^{-1})$ 	&		&		&		& Obs\,2: $<0.32$ \\	
        Obs\,1 Photon index ($\Gamma$)	&	2.2	&	$[2.09, 2.26]$	&	$[2.02, 2.29]$	& $[1.88,1.96]$\\	
        Obs\,1  $L_{2-10}/L_{\rm Edd}$ ($\%$)	&	0.25	&	$[0.23,0.29]$	&	$[0.22,0.31]$	& $[0.5,0.6]$ \\	
        Obs\,2 photon index ($\Gamma$)	&	2	&	$[1.90,2.05]$	&	$[1.87,2.08]$	& $[1.99,2.07]$\\	
        Obs\,2  $L_{2-10}/L_{\rm Edd}$ ($\%$)	&	0.19	&	$[0.18,0.21]$	&	$[0.17,0.23]$	& $[0.4,0.5]$\\	
        Soft power-law fraction $C_{\rm PL}$ ($\%$)	&	2.4	&	$[1.8,3.0]$	&	$[1.4,4.9]$	& N/A\\	
        Soft power-law  $\Pi_{\rm PL} \,(\%)$	&	55	&	$[21,95]$	&	$>7$	& N/A\\	
        Soft power-law $\Psi_{\rm PL} \,(\degr)$	&	16	&	$[-2.7,34]$	&	$[-12,50]$	& N/A \\ 
    \hline	
    \end{tabular}
    \caption{Best-fit parameters obtained by modeling the IXPE Stokes parameters (columns 2-4), and the \xmm\ and \nustar\ spectra (last column).}
    \label{tab:Polpar}
\end{table*}

The left column of Fig.\,\ref{fig:polvsenergy} shows the resulting best-fit polarization fraction and polarization angle (top and bottom panels, respectively) as a function of energy from the full model. In the right column, we show the polarization properties by isolating each of the model components. We remind the reader that the polarization fraction and angle are not additive quantities. However, this component separation provides an informative idea on how each of the components compare to the observed data, and the level of specific polarization expected from each component when treated separately.

The red solid lines correspond the total polarization properties obtained from the disk reflection and the primary unpolarized coronal emission. It is clear that this component alone can account for the observed polarization fraction. Isolating the disk reflection alone (grey dotted line), we find that this component can reach $\pdx \simeq 16\%$. This is broadly consistent with the approximated value of the reflection component reported by \cite{Gianolli2024}. Interestingly, the predicted \pax\ from the disk reflection is $\sim 91-95^\circ$, providing a good description of the values measured by \ixpe\ above 4\,keV. Adding the unpolarized corona emission does not affect the polarization angle.

The cyan dashed lines and the pink dash-dotted lines show the polarization parameters of from the torus and the soft PL, respectively. If these components were considered alone, their respective polarization fractions may exceed that of the disk reflection \pdx. But given their low flux compared to the corona and disk reflection, their effects are minimal at hard X-rays. In the soft X-rays, as the corona and disk flux gets dimmer due to absorption, the contribution of these components is more effective. Their effect is more noticeable in the soft X-rays, as their corresponding polarization angles are lower than that of the disk reflection which will lead to a larger change in the total polarization (when all components are added together) below $\sim 4\,$keV. This effect is more important for the soft PL component which has a larger \pdx.

To estimate the uncertainty on the main parameters, we froze the absorbers column densities to their best-fit values ($N_{\rm H, low} = 5 \times 10^{22}\,\rm cm^{-2}$ and  $N_{\rm H, high} = 9.2/7.8 \times 10^{22}\,\rm cm^{-2}$ for Obs\,1/2). Then we calculated the 1D contour plots (using XSPEC's \texttt{steppar} command) for each of the remaining parameters as shown in Fig.\,\ref{fig:chi2_parameters}. The uncertainties on these parameters are reported in Table\,\ref{tab:Polpar}. The blue horizontal lines in this figure correspond to the $1\sigma, 2\sigma,$ and $3\sigma$ ($\Delta \chi^2 = 9.3, 15.8, 23.6$, respectively) confidence levels for 8 free parameters. We do not count $\Gamma$ and $L_{2-10}/L_{\rm Edd}$ for each observations twice, otherwise the $\Delta \chi^2$ margins will be even higher than the ones used. This figure shows that modeling the polarization signal in this source assuming a lamp-post geometry requires a relatively low coronal height of $h < 9\,r_{\rm g}$ (at $3\sigma$). We also found that the lower limit of the torus opening angle is $45\degr$ (at $3\sigma$). The additional component that is required to model the polarization below $\sim 4$\,keV consists of a small fraction of the coronal primary flux (between 1 and 5\%, at $3\sigma$). We found that the polarization fraction of this component to be larger than 10\% (at $3\sigma$), and its polarization angle is found to be $\sim 16\degr$.

\section{The energy spectrum}
\label{sec:spectra}

\subsection{IXPE}
\label{sec:ixpespec}

\begin{figure*}
\centering
\includegraphics[width=0.99\linewidth]{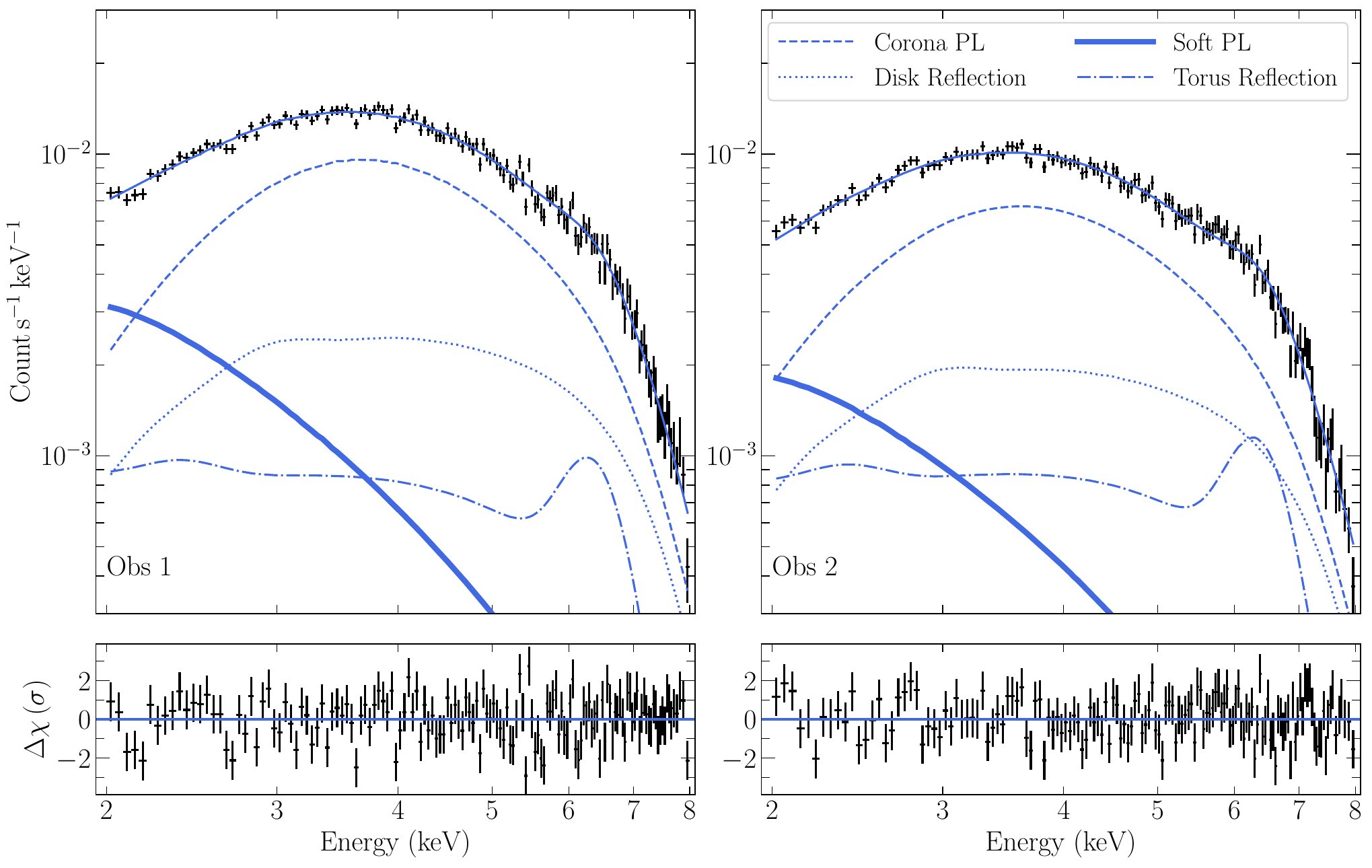}

\caption{Time-integrated \ixpe\ energy spectra obtained from Obs\,1 (left) and Obs\,2 (right) are shown in the top panel. We show the best-fit model as described in Sect.\,\ref{sec:ixpespec}. The dashed lines correspond to the coronal power-law component. The dotted lines correspond to the disk reflection. The dash-dotted lines correspond to the torus emission. The thick solid lines correspond to the additional power-law component required to model the polarization signal below 4\,keV. The corresponding residuals are shown in the bottom panel.}
\label{fig:ixpe_spec} 
\end{figure*}

The spectro-polarimetric fits presented in the previous section were performed on the Stokes parameters ($I, Q, U$) binned into five broad energy bins, in order to maximize the polarization signal-to-noise ratio. In this section, we test whether the same best-fit model is also consistent with the full-resolution IXPE energy spectrum (i.e., the Stokes $I$ parameter alone), which provides a more stringent check on the spectral shape of the individual model components.

We considered the total energy spectrum from each of the three \ixpe\ detectors for both observations. We applied the same model described in the previous section. To test the consistency between the results from the previous section and the energy spectrum, we allowed the parameters to vary only within the $1\sigma$ limits shown in Fig.\,\ref{fig:chi2_parameters}. We show the combined energy spectrum from the three detectors in Fig.\,\ref{fig:ixpe_spec}  for Obs\,1 (left) and Obs\,2 (right). This figure shows that the best-fit model that explains the polarization signal is able to describe well the time-integrated \ixpe\ spectrum in both observations with a total $\chi^2/\rm dof = 970/875$ ($p_{\rm null}=0.014$). In addition, this figure shows that the coronal power-law emission is the dominant component. While the disk reflection component contributes to $\sim 22-23\%$ of the total flux in the $2-8$\,keV range probed by \ixpe.

\subsection{XMM-Newton and NuSTAR}
\label{sec:xmm}

As a final test of the viability of the X-ray reflection hypothesis, we applied our model to the \xmm\ and \nustar\ spectra, over the $0.4-79\,\rm keV$ range, taken during the \ixpe\ observations. We stress that the goal of this exercise is to test whether the model used earlier to fit the X-ray polarization data can also explain the broadband spectra with consistent parameters. We do not aim to perform a detailed spectral analysis of the \xmm\ and \nustar\ observations.

\begin{figure*}
\centering
\includegraphics[width=0.99\linewidth]{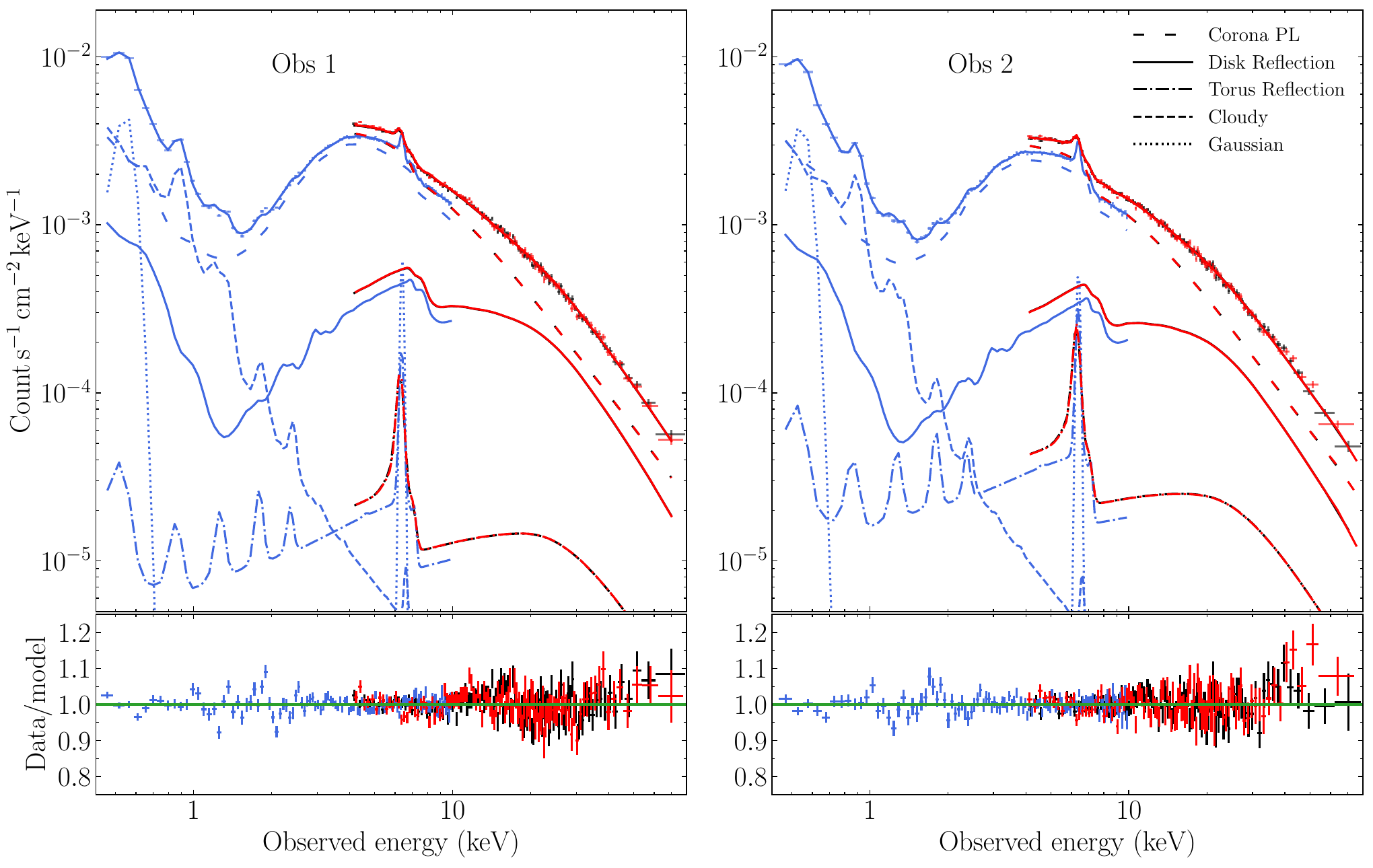}

\caption{Best-fit model to the XMM-Newton (blue) and NuSTAR/FPMA and FPMB (black and red, respectively) assuming disk reflection for Obs\,1 and Obs\,2 (left and right, respectively). The spectra are re-binned for visual purposes.}
\label{fig:xmm_nustarspec}
\end{figure*}


We replaced {\tt KYNSTOKES} by {\tt KYNSED}\footnote{\url{https://projects.asu.cas.cz/dovciak/kynsed}} \citep{Dovciak2022}, which is more suitable for fitting the higher energy resolution spectra. Both models account for all relativistic effects within the framework of a point-like X-ray corona illuminating the accretion disk. Since the energy range covered by \xmm\ and \nustar\ is wider than that of \ixpe, it is necessary to consider a more complex treatment of absorption and emission in the soft X-ray band. We added a Galactic absorption \citep[\texttt{TBabs;}][]{wilms00} fixed at $N_{\rm H} = 2.07\times 10^{20}\,\rm cm^{-2}$ \citep{HI4PI}. Similarly to \cite{Gianolli23, Gianolli2024}, we used two neutral partial covering absorbers ({\tt zpcfabs}) and an ionized absorber ({\tt zxipcf}). To model the soft X-ray emission, we added the same photoionized plasma emission component used by \cite{Gianolli23, Gianolli2024}, produced with \textsc{CLOUDY} \citep{Ferland1998, Gunasekera2025cloudy}. We note that the different shape with respect to the soft PL used to model the \ixpe\ data is not very significant above $\sim3$\,keV, since the flux of the component is very low. We also added a Gaussian emission line at $\sim 0.5\,\rm keV$ to account for the \ion{O}{7} emission-line triplet, which is not perfectly captured by the photionization model. For the neutral reflection we used {\tt stokes\_torus}. This model includes a narrow Fe\,K$\alpha$ line at $\sim 6.4$\,keV, in addition to other lines in the soft X-rays. We also added a Gaussian emission line fixed at 6.4\,keV to account for the broad-line-region-like line discovered in this source using the XRISM/Resolve high-resolution spectra \citep{Xrism2024ngc4151}. We let the normalization of the line to be free with an unresolved line width fixed at zero. In addition, the data required the addition of four absorption lines at $\sim 6.7$\,keV, 6.9\,keV, 7.4\,keV, and 7.75\,keV. These lines are consistent with the ones found in the XRISM/Resolve spectra of the source as reported by \cite{Xiang2025}. We added a cross-normalization constant\footnote{A cross-calibration difference of $\sim 18-20\%$ is already expected between XMM-Newton and NuSTAR as detailed in the XMM-Newton release note \href{https://xmmweb.esac.esa.int/docs/documents/CAL-SRN-0388-1-4.pdf}{XMM-CCF-REL-388}.} between \xmm\ and \nustar\ of the order of $1.2$. The model can be expressed in \textsc{Xspec} parlance as follows:

$${\tt model = Constant \times  TBabs[  }$$
$$ \tt{zpcfabs \times zpcfabs \times zxipcf \times  KYNSED\,+}$$
$${\tt  stokes\_torus + Gaussian_{6.4\,keV} + CLOUDY +}$$
$${\tt Gaussian_{0.5\,keV}  + 4\times Gaussian_{abs}} ].$$

As before, we fixed the inclination of the system to 60\degr\ for all the model components that require this parameter. We left the BH spin, the corona height, and the photon index free to vary for {\tt KYNSED}. {\tt KYNSED} parametrizes the X-ray luminosity using the ratio of power transferred to the corona to the total accretion luminosity of a Novikov-Thorne disk \citep{Novikov73}, referred to as $L_{\rm transf}/L_{\rm disk}$. If $L_{\rm transf}/L_{\rm disk}$ is positive, this parameter represents the power that is transferred from the disk to the corona from below a transition radius $r_{\rm trans}$. If  $L_{\rm transf}/L_{\rm disk}$ is negative, the model assumes that the power is given to the corona from an external source. In this case, the transition radius is set at the ISCO and $L_{\rm transf}/L_{\rm disk}$ can take any negative value. We refer the reader to \cite{Dovciak2022} for more details of this parameter. In this work, we use $L_{\rm transf}/L_{\rm disk} <0$, which is consistent with the geometry used by \texttt{KYNSTOKES} to model the \ixpe\ data. Once a fit is performed, the \texttt{xset} command in XSPEC can be used to obtain the total X-ray luminosity. We used this value to estimate the $L_{2-10}/L_{\rm Edd}$ for the best-fit $\Gamma$ and compare the results with the \ixpe\ values.

For {\tt stokes\_torus}, we allowed the opening angle and the normalization to vary within the corresponding $3\sigma$ confidence ranges that we obtained by modeling the \ixpe\ data. We note that the soft power-law component used to model the \ixpe\ polarization data and that consists of $\sim 2.5\%$ of the primary power-law flux is not included in the model of the \xmm\ and \nustar\ spectra. The complexity of the model we use in the soft X-rays does not allow to identify such a faint component in the data. In fact, the \textsc{cloudy} model component could substitute the soft PL, as it has a continuum component which can be approximated by a power law. Finally, given the high quality of the data and the limitations of any model to capture all the details in the spectra, we added a gain shift to each of the instruments that turned out to be of the order of $\sim 2-5$\,eV for \xmm\ and $\sim -60$\,eV for \nustar\ (comparable to the channel width), consistent with the value found by \cite{Gianolli23}. 

The spectra and the best-fit models are shown in Fig.\,\ref{fig:xmm_nustarspec}. The model results in fits with $\rm\chi^2/dof = 715/519$ and $ 597/498$ for Obs\,1 and Obs\,2, respectively.  Formally, these fits are not statistically acceptable. The large $\rm\chi^2$ values are mainly driven by the fit to the soft X-rays. This is evident from the data/model ratios shown in the lower panels of Figure \ref{fig:xmm_nustarspec}. These plots suggest that the main discrepancy between the data and the best-fit model is due to narrow features in the $\sim 0.8-2.5$ keV band, which cannot be taken into account by the model. In fact, if we ignore all the data below 3\,keV, we obtain acceptable fits with $\chi^2/{\rm dof} = 563/474$ ($p_{\rm null}=0.003)$ and $519/454$ ($p_{\rm null}=0.02$) for Obs\,1 and Obs\,2, respectively. 

If we consider only the data below 3\,keV, we obtain $\chi^2/{\rm dof} = 152/21$ and $79/20$ for Obs\,1 and Obs\,2, respectively. The average data/model ratio in this energy band is of the order of a few per cent. We can get an estimate of the mean amplitude of the residuals in the soft band by introducing a systematic error, so that the fit becomes acceptable.  We gradually increased the systematic error that we added in the model (in quadrature) until we obtained an acceptable fit below 3 \, keV with $p_{\rm null}=0.01$. Systematic errors of $\sim 2.9$\% and $2.1$\% are then required for Obs\,1 and Obs\,2, respectively. We therefore conclude that the model fits the energy spectra well at energies higher than 3\,keV. At lower energies, there are discrepancies between the best-fit model and the data with an average amplitude of $\sim 2-3$\%, on average, which are due to narrow features that cannot be explained by the model.

In order to calculate the error on the best fit parameters, it is necessary to work with a model that provides acceptable fits to the data. To this end, we added a systematic error of 1.5\% to the model for both observations. This reduces $\rm\chi^2/dof$ to = $511/519$ and $ 483/498$ for Obs\,1 and Obs\,2, respectively, without affecting any of the best-fit parameters or the distributions of the data-to-model ratios. The 3$\sigma$ confidence regions listed in Table\,\ref{tab:Polpar} were calculated this way. We stress that we do not claim that the model provides a good fit to the spectra in the full energy band. We simply increase the systematic error so that we can estimate a reasonable confidence region for the best-fit parameters. If we do not do it this way, the resulting confidence regions will be too small, and perhaps even meaningless.

During the fitting, we left the spin parameter free to vary. Unlike with the \ixpe\ data, where some constraints on the BH spin could be obtained, we were unable to constrain the spin parameter when fitting the \xmm\ and \nustar\ data. Nevertheless, the inferred values are consistent within $3\sigma$ with the values obtained from modeling the \ixpe\ data.

The best-fit parameters are shown as vertical lines in Fig.\,\ref{fig:chi2_parameters}. We found coronal heights (with a 3$\sigma$ range) of 5.2\,\rg\ ($<9\, \rg$) and 3\,\rg\ ($< 4.5$\,\rg) for Obs\,1 and Obs\,2, respectively. These values are fully consistent with the value inferred from modeling the \ixpe\ data. We find photon indices of 1.95 and 2.03 for Obs\,1 and Obs\,2, respectively. 

Our model results in 17\% and 18\% contributions of the disk reflection to the total observed flux in the $2-8$\,keV range in Obs\,1 and Obs\,2, respectively. These values are compatible with the disk reflection contribution estimated by modeling the \ixpe\ data.

As for the distant, torus reflection, we were able to infer only upper/lower limits on the normalization/opening angle, which are consistent with the values inferred from modeling the \ixpe\ data. In fact, we expect to have some level of degeneracy between the torus reflection and the BLR-like emission line at 6.4\,keV. We also found that the torus reflection emission varies between the two observations, where the torus contribution is smaller in Obs\,1. This is in agreement with the fit of the IXPE spectra shown in Fig.\,\ref{fig:ixpe_spec}. 

In addition, we left the high-energy cutoff free throughout the fitting. We found a value of 300\,keV in Obs\,1. We were not able to constrain the cutoff for the second observation. It is worth noting that, although the fits are statistically acceptable at energies higher than 3 keV, a small excess can be seen at energies above $\sim 40-50$\,keV in the residual plots of Fig.\,\ref{fig:xmm_nustarspec}. These residuals could be due to a variety of assumptions and factors like the treatment of Comptonization in the reflection models, or the treatment of the high-energy cutoff. Exploring these effects is beyond the scope of this work. 

The best-fit neutral absorption parameters we obtained for Obs\,1 (Obs\,2) are $N_{\rm H} = 1.3 (6) \times 10^{23}\,\rm cm^{-2} $ and $f_{\rm cov} = 0.6 (0.3)$ for the first absorber, and $N_{\rm H} = 5.8 (7.4) \times 10^{22}\,\rm cm^{-2} $ and $f_{\rm cov} = 0.94 (0.75)$ for the second absorber. As for the ionized absorber we obtained $N_{\rm H} = 1.7 (7.5) \times 10^{22}\,\rm cm^{-2} $, $\log (\xi/{\rm erg\,cm\,s^{-1}}) = 1.3 (1.1)$, and $f_{\rm cov} = 0.81 (0.97)$. We note that we fix these parameters to their best-fit values when we estimated the uncertainties on the other parameters reported in Table\,\ref{tab:Polpar}.

It should be noted that a direct 1:1 comparison between the spectral fits presented by \cite{Gianolli23, Gianolli2024} and our fits cannot be performed, as different GTIs and binning schemes are used for Obs\,1, while our \xmm\ and \nustar\ Obs\,2 was not analyzed by \cite{Gianolli2024}. Thus, for reference, we fitted the same model used by \cite{Gianolli23, Gianolli2024} to the spectra presented in this work. This results in $\chi^2/{\rm dof}= 756/521$ and $682/500$, for Obs\,1 and Obs\,2, respectively, when fitting the full energy band spectra. Thus, neither the \cite{Gianolli23, Gianolli2024} model can fit the combined \xmm\ and \nustar\ well. Formally speaking, our model reduces the value of $\chi^2_{\rm min}$ by 41 and 85 (for two additional degrees of freedom) for the spectra of Obs\,1 and Obs\,2. However, this comparison may not be important, given the fact that neither of the two models can fit the data well (from a statistical point of view).

\section{Discussion and Conclusions}
\label{sec:conclusion}

We have presented an X-ray spectro-polarimetric analysis of the \ixpe\ observations of \ngc. This is the only non-jetted type-1 AGN where an X-ray polarization signal has been detected with more than 3$\sigma$ significance, reaching $\sim 6-7\%$ above 4\,keV. The polarization angle inferred from the \ixpe\ data suggests the presence of two components contributing to the polarization signal: a hard component above 4\,keV with a polarization angle parallel to the jet in this source, and a softer component with a different polarization angle affecting the softer X-rays. 

We demonstrate that the disk reprocessing scenario, within the lamp-post corona framework, is indeed capable of explaining the polarization signal above 4\,keV. We found an additional component affecting the soft X-ray polarization with $\pdx > 7\%$ (at 3$\sigma$) and $\pax \sim 16\degr$, consistent with the findings of \cite{Gianolli2024}.

\ngc\ is the only non-jetted and mildly absorbed AGN with a detected X-ray polarization, which enables studying the geometry of the accretion flow around black holes. A different example is the X-ray binary system Cygnus\,X-1 observed by IXPE in the hard state \citep{Krawczynski2022, Kravtsov2025}. The polarization properties could be explained via Comptonization in equatorially extended coron\ae\ for higher inclinations ($\gtrsim 45^\circ$) than inferred from optical observations ($\sim 30^\circ$), unless a fast outflow of the corona plays a role \citep{Krawczynski2022, Poutanen2023}. Several configurations examined with {\tt KYNSTOKES} code in the lamp-post geometry for unpolarized corona, including different examples of disk truncation expected in the hard states of X-ray binaries, did not lead to a satisfactory model match with the observed X-ray relativistic reflection fraction in $2-8$\,keV ($\sim10\%$) and the relatively high $\pdx \sim4\%$ at the same time for any inclination \citep{Krawczynski2022}. The disk-reflection origin of X-ray polarization with a vertically extended corona, if eventually preferred over the equatorially extended coronal origin for \ngc\ and similar sources, may also be a feature of AGNs not inherent to the X-ray binary systems in their hard state. However, that can only be tested on larger samples and with higher quality of X-ray spectro-polarimetric data. A detailed test of the disk reflection model, similar to the one presented in this work, applied to Cyg X-1 in its different states will be important to understand the geometry of the corona around accreting black holes.

\subsection{Comparison of the polarization and spectroscopy results}

In summary, we find good agreement between the spectro-polarization analysis using \ixpe\ and the spectral analysis of the \xmm\ and \nustar\ observations. Both data sets could be well described by assuming a geometry in which the accretion disk is illuminated by a centrally located X-ray source. The X-ray illumination model fits well the \xmm\ and \nustar\ spectra at energies above 3 keV, but there are significant residuals at lower energies, albeit of small amplitude on the order of $\sim 2-3$\%. These discrepancies are due to narrow features in the $\sim 0.8-2.5$\,keV range, and to some extent calibration uncertainties in the \xmm\ PN detector may be responsible for these features in the residual plots.

The good agreement between the best-fit results when we fit the \ixpe\ spectro-polarimetric data and the \xmm\ and \nustar\ spectra is even more impressive if we consider that, the disk reflection models used to fit the \ixpe\ (\texttt{KYNSTOKES}) and the \xmm\ and \nustar\ spectra (\texttt{KYNSED}) assume similar geometries and physics, they also differ especially in their treatment of the ionized reflection. \texttt{KYNSTOKES} is based on calculations by \cite{Podgorny22} while \texttt{KYNSED} is based on the \texttt{XILLVER} tables by \cite{Garcia2013, Garcia16}. The latter model is more appropriate for fitting the energy spectra of \xmm\ and \nustar\ due to its higher spectral resolution. These differences could lead to small changes in the inferred parameters. 

In addition, \ngc\ is known to be highly variable, as shown in Fig.\,\ref{fig:lcfig}. This figure shows that the \xmm\ and \nustar\ observations capture the source in states that are $\sim 25$\% and 40\% brighter than the average flux in Obs\,1 and Obs\,2, respectively. This would also lead to variability in the physical parameters, notably the coronal height and photon index \citep[see][]{Kammoun2024, Papoutsis2026}. For example, the difference in the source flux between the \ixpe\ data and the \xmm\ and \nustar\ observations may explain the difference we detect between the best-fit $\Gamma$ values. AGN show spectral slope variations, which are often associated with the source flux in the $2-10$\,keV band \citep[e.g.,][]{Sobolewska2008}. In Obs\,1, the \ixpe\ photon index is steeper than the one of the Obs\,2. This is consistent with the fact that the source flux in the second observation was lower. The best-fit photon index of the second \xmm\ and \nustar\ observation is broadly consistent with the \ixpe\ Obs\,1 best-fitted value. This is also expected as the source flux is similar in both observations (see Fig. \ref{fig:lcfig}). It is only the best-fit spectral slope of the first \xmm\ and \nustar\ observation that does not agree with the ``softer when brighter'' trend that is usually observed in Seyfert galaxies, but this trend also shows a considerable scatter \citep[see e.g.,][]{Sobolewska2008}.

Despite these caveats, we find good agreement in most physical parameters. In fact, the height inferred from the \xmm\ and \nustar\ observations is found to be $<9\,\rg$ and $<4.5\,\rg$ (at $3\sigma$) in Obs\,1 and Obs\,2, respectively. This is fully consistent with the $3\sigma$ upper limit of $9\,\rg$ obtained from modeling the \ixpe\ data. This value is consistent with the expectation from X-ray spectral-timing analyses of nearby bright AGN \citep[e.g.,][]{Kara2016, Caballero18, Caballero2020}. 

From the \xmm\ and \nustar\ observations, we could derive only lower limits on the torus opening angle and upper limits on the normalization of the torus reflection. These limits are consistent with those obtained from modeling the \ixpe\ data. It is worth noting that we expect these two parameters to be degenerate. In addition, as our global model consists of three iron lines in total (disk reflection, BLR, and torus), consistent with the XRISM results, an additional level of degeneracy can be expected notably between the BLR and the torus lines.

\begin{figure*}
\centering
\includegraphics[width=\textwidth]{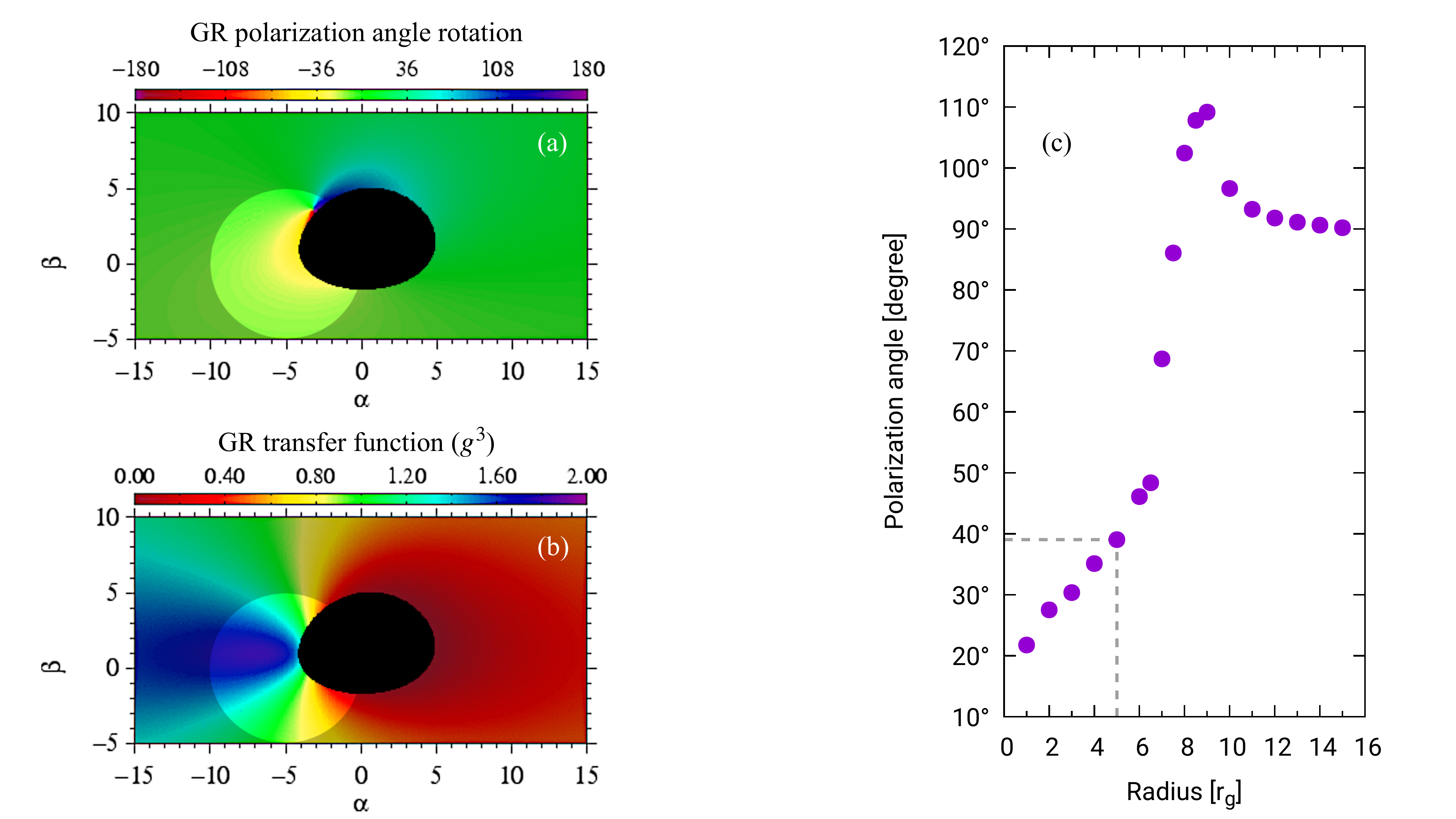}
\caption{a) GR rotation of the polarization angle (in degrees; denoted by the color scale) for photons originating from different parts of the inner accretion disk residing in the equatorial plane, as projected onto the observer's sky (with impact coordinates $\alpha$ and $\beta$ in units of $r_{\rm g}$). The black region denotes the area below the ISCO. The bright circle marks a region of lower absorption. The BH spin and inclination were assumed as in our fits, i.e., $a^*=0.75$, $i=60^\circ$. b) Same as a), but showing the GR transfer function given by the cubed g-factor (energy shift), $g^3$, with the color scale representing flux amplification due to GR effects. One can see that the brightest regions correspond to a particular GR rotation of the polarization angle. c) Predicted polarization angle, \pax, for different sizes of the less obscured region shown in the left panels, centered at $\alpha=-5\,r_{\rm g}$ and $\beta=0$. The {\tt KYNSTOKES} model with the best-fit parameter values from the first observation was used, accounting only for emission from the less obscured part of the disk. Note that the final polarization angle is computed from the local polarization angles at the emission points, the GR rotations of the polarization angles, and the position angle of the system axis. The observed \pax\/ corresponds to a size of approximately $5\,r_{\rm g}$ for the less-obscured region. However, contributions from more absorbed parts of the disk must also be taken into account to properly estimate this size.}
\label{fig:GR-PA-rotation}
\end{figure*}

\subsection{The origin of the soft X-ray polarization}
\label{sec:softPLorigin}

As discussed in Section\,\ref{sec:ixpe}, a polarized soft power-law component, whose physical origin is not specified by the fit, is needed to explain the polarization signal below 4\,keV. This component accounts for only a few percent of the primary power-law flux. Thus, given the complexity of the spectral features in the soft X-rays, this component is not required when modeling the energy spectra. However, it could be well hidden under either the \texttt{CLOUDY} component or taken into account by the partial covering absorption. As Fig.\,\ref{fig:polvsenergy} and Fig.\,\ref{fig:chi2_parameters} show, the polarization angle of this component is different from all the other components, which drives the polarization signal in the soft X-rays. The circumnuclear region in \ngc\ is known to be complex. Multi-wavelength observations of the inner $\sim 10\,\rm pc$ show the presence of dusty outflowing structures \citep[see e.g.,][]{Kishimoto2022, Isbell2026}, which could explain the presence of this additional polarized component due to scattering.

Here, we present two alternative interpretations that do not require a separate scattering region, but instead attribute the soft polarization signal to effects already operating on the disk and coronal emission in our model. First, partial covering of the innermost regions of the accretion flow could result in an effect similar to the one we see in the soft X-rays. In fact, the analysis of the broad-band X-ray spectra of the source require the presence of partial covering absorption. This could be explained by clumpy structure in the BLR and/or the torus. \cite{Kammoun2018eclipses} studied the effect of X-ray eclipses on the polarization signal in AGN. They showed that as a BLR cloud passes through the line-of-sight, different regions of the accretion disk will be covered/uncovered, which affects the observed polarization fraction and angle. Consequently, a similar effect could be causing the change in polarization angle at soft X-rays in \ngc\ due to emission from different parts of the accretion disk. In fact, absorption is more effective at soft X-rays rather than at hard X-rays where the photons may pass through the covering cloud, which preserves \pax. Therefore, while at higher energy, the polarization angle is determined by summation of the Stokes parameters over the whole accretion disk illuminated by the corona, at low energy the emission from the least absorbed parts of the disk will prevail. In Fig.\,\ref{fig:GR-PA-rotation}, we show an example where the approaching, Doppler-boosted part of the accretion disk, where the disk is the brightest (panel b), is the least absorbed (denoted by the bright circular region). At the same time, in this part of the disk, the GR rotation of the polarization angle reaches up to $40^\circ$ clockwise from the system axis (panel a), which is consistent with the rotation observed between high and low energies in the IXPE observation of \ngc. The final observed polarization angle then depends on the position and size (panel c) of the less absorbed region as well as the overall contribution from other more absorbed regions of the accretion disk.

Alternatively, the behavior seen in soft X-rays could be caused by the steep energy dependence of photo-electric absorption opacity. This dependence causes a change in the X-ray polarization properties in the soft band with partially obscured emission, as opposed to the hard band where the absorber becomes transparent. \cite{Marin2018b} and \cite{Podgorny2023torus,Podgorny2024} show the X-ray polarization dependence on the column density and geometry of a neutral obscurer and the inclination of the observer. At soft X-rays the polarization degree and angle are different from the hard X-rays with a relatively sharp energy transition between parallel and perpendicular polarization angle for a symmetrically obscured axially symmetric source. \cite{VanderMeulen2024} also show the X-ray transition energy between obscured and unobscured (soft and hard) polarization properties. Particularly, their Fig.\,9 (middle left panel) is illustrative of an energy transition near 3.5\,keV, similar to the one detected in \ngc. In our interpretation, the absorber does not need to be fully neutral and we do not presume a symmetric obscuration. Hence, a generally different soft X-ray polarization angle than parallel or perpendicular to the jet direction is expected, as well as different steepness of the polarization energy transition. Further X-ray polarimetric observation of \ngc\ may catch the partial absorber in a different position relative to the source. Changes in the $2-3$\,keV polarization degree and angle values and consistency of the  $4-8$\,keV polarization state may thus confirm such interpretation.

Finally, we consider whether the radio jet itself contributes to the observed polarization. Jet-aligned X-ray polarization is established in beamed radio-loud AGN, where it arises from synchrotron emission within the jet \citep[see e.g.,][and references therein]{Marscher2024, Capecchiacci2025}. We stress that this is a different physical mechanism from the one considered here. In our model the polarization is aligned with the jet because disk reflection polarizes the emission along the disk symmetry axis, which is expected to be roughly parallel to the jet since the jet is launched perpendicular to the disk. A jet-synchrotron origin is, in any case, not viable for the hard-band signal in \ngc. A beamed, pole-on jet is excluded by the source itself. \ngc\ is radio-quiet \citep[$\log(L_{\rm R}/L_{\rm X})<-5$;][]{Ulvestad2005}, and its narrow Fe\,K$\alpha$ line and optical broad-line region are characteristic of an accretion-powered nucleus rather than a Doppler-boosted continuum. In addition, its milli-arcsecond radio nucleus is faint ($\sim 10^{37}\,\rm erg\,s^{-1}$) and two-sided \citep{Ulvestad2005}, unlike a jet beamed toward us. A non-beamed jet is excluded by the polarization geometry. \citet{Lyutikov2005} show that jet-parallel synchrotron polarization is geometrically restrictive, arising only for jets viewed close to their axis ($\theta \lesssim 1/\Gamma$) and requiring a rest-frame field dominated by its toroidal component. At the viewing angles relevant here ($\theta \approx 45-90^{\circ}$, for a jet close to the plane of the sky and a system inclination of $60^{\circ}$), a relativistic synchrotron jet ($\Gamma \sim 10$) is instead polarized either orthogonal to the jet or negligibly, at all magnetic pitch angles (their Fig.\,7), and therefore cannot reproduce the observed jet-parallel signal. The hard-band polarization is thus well described by disk reflection alone. A sub-dominant jet contribution cannot be excluded, but by construction it cannot account for the observed $\sim6-7\%$ signal.

It is worth noting that the geometric interpretation of the X-ray polarization relies on the assumption that the extended radio emission traces a jet whose orientation can be extrapolated to sub-pc scales and is orthogonal to the accretion flow. Neither condition is guaranteed: the radio position angle in \ngc\ varies with angular resolution \citep{Mundell2003}, and misalignment between the jet and disk axes have been suggested \citep{May2020}. Moreover, in nearby Seyfert galaxies the radio emission may originate from wind-driven shocks rather than a jet \citep[e.g.,][]{Fischer2023}, in which case the coincidence between the radio and X-ray polarization angles would carry no geometric information about the corona, substantially weakening the constraints on its shape. We stress that this assumption is not merely a supporting argument but a necessary condition of the interpretation. The polarization angle predicted by disk reflection is aligned with the projected disk symmetry axis, whose orientation on the sky is fixed, in our fits, by the adopted position angle of $83^{\circ}$. Should the inner accretion flow, and hence the inner jet, be substantially misaligned with the resolved radio structure, the appropriate reference direction would no longer be the observed radio position angle, and the model would no longer reproduce the polarization angle measured above 4\,keV. The same holds for the corona-dominated interpretation of \citet{Gianolli23, Gianolli2024}, which predicts a polarization angle aligned with the same system axis. This is therefore a limitation shared by all geometric interpretations of the X-ray polarization in this source, rather than one specific to the disk-reflection scenario. We note, finally, that the coincidence of the hard-band polarization angle with the radio position angle is itself an argument against a strong misalignment, since an inner axis oriented differently would generically produce a polarization angle displaced from the jet direction. Higher-resolution constraints on the inner jet orientation, or X-ray polarimetry in different spectral states, would help to test this assumption directly.

\subsection{Comparison with the corona-only scenario}
\label{sec:comparison}

\cite{Gianolli23, Gianolli2024} suggested that X-ray polarization is induced directly by Comptonization in a corona with an equatorially extended geometry. The existence of both solutions (extended corona and disk reflection) is well understandable from the parametric examination of the underlying model output. For an unpolarized point-source corona, at a relatively low height ($\lesssim 20 \,\rm \rg$), irradiating a partially ionized disk, the expected behavior of mid-X--ray \pdx\ and \pax\ as a function of inclination is similar to an equatorially extended corona (cf., e.g., Figs.\,$6-7$ of \citealt{Podgorny23kyntokes} and Figs\,$4-13$ of \cite{Tagliacozzo2025} or Fig.\,2 of \citealt{Poutanen2023}). In the case of a polarization from X-ray reflection, \pdx\ increases from 0 to $\lesssim 20\%$ from a face-on to an edge-on configuration, while \pax\ is nearly aligned with the projected system axis of symmetry for all inclinations.  Thus, the dependence of \pdx\ on inclination, in the case of X-ray illuminated disks, may naturally explain the lack of polarization detections in other Seyfert\,1 galaxies observed by \ixpe, all of which have lower estimated inclinations than \ngc.

\subsection{Conclusions}

We studied in detail the two \ixpe\ observations of \ngc, together with contemporaneous \xmm\ and \nustar\ observations. We found that X-ray reflection from the accretion disk can fit the data with a BH spin of 0.75, an inclination angle of 60\degr, and accretion rate of 0.02 of the Eddington limit, and an X-ray corona height smaller than $\sim 9\,\rg$.  The model can fit both the X-ray polarization properties (above 4\,keV) and the \xmm\ and \nustar\ energy spectrum above 3\,keV. At lower energies, we detect significant residuals in the \xmm\ spectra, but their amplitude is small (of the order of $\sim 2-3\%$). These discrepancies could be due, to some extent, to calibration uncertainties of \xmm\ at soft energies. New \ixpe\ observations of \ngc\ in different spectral states (particularly states in which the reflection fraction is either dominant or negligible) together with observations of additional high-inclination Seyfert\,1 galaxies will provide the most stringent test of the disk reflection scenario presented in this work.

Below 4\,keV, the \ixpe\ data require an additional polarized component carrying only $\sim 1-5\%$ of the primary coronal flux but with a polarization degree $> 10\%$ (at $3\sigma$) and a polarization angle of $\sim 16\degr$, distinct from those of the disk reflection and the torus. In Section~\ref{sec:softPLorigin}, we discussed three possible origins for this soft component: scattering in the dusty, parsec-scale structures revealed by recent mid-infrared interferometry of \ngc, 
as also suggested by \cite{Gianolli2024}; partial covering of the innermost accretion flow by clumpy BLR or torus material, which can preferentially obscure regions of the disk where the local GR rotation of the polarization angle reaches up to$ \sim 40\degr$ from the system axis (Fig.~\ref{fig:GR-PA-rotation}); or a transition in the polarization properties near $\sim 3$\,keV driven by the steep energy dependence of the photoelectric opacity in a non-neutral absorber. 

\facilities{IXPE, XMM, NuSTAR}

\begin{acknowledgements}
We would like to thank the anonymous referee for their feedback. E.K. would like to thank Giorgio Matt and Travis C. Fischer for useful discussions.   E.K., S.B., F.U. acknowledge financial support by the Italian Space Agency (Agenzia Spaziale Italiana, ASI) through the contract ASI-INAF-2022-19-HH.0. M.D. and J.P. thank GACR project 26-22614S. M.D. and J.P. acknowledge institutional support from RVO:67985815. V.B.-V. acknowledges funding by the European Union ERC-2022-STG-BOOTES-101076343. V.E.G. acknowledges funding under NASA contract 80NSSC24K1403. 

This research has made use of data and/or software provided by the High Energy Astrophysics Science Archive Research Center (HEASARC), which is a service of the Astrophysics Science Division at NASA/GSFC and the High Energy Astrophysics Division of the Smithsonian Astrophysical Observatory. This research made use of XSPEC \citep{Arnaud1996}. This research made use of SciPy \citep{Virtanen_2020}, matplotlib, a Python library for publication quality graphics \citep{Hunter:2007} and NumPy \citep{harris2020array}. This research is based on observations obtained with the Imaging X-ray Polarimetry Explorer (IXPE), a joint US (NASA) and Italian (ASI) mission, led by Marshall Space Flight Center (MSFC). The research uses data products provided by the IXPE Science Operations Center (MSFC), using algorithms developed by the IXPE Collaboration (MSFC, Istituto Nazionale di Astrofisica - INAF, Istituto Nazionale di Fisica Nucleare - INFN, ASI Space Science Data Center - SSDC), and distributed by the High-Energy Astrophysics Science Archive Research Center (HEASARC).  This research is also based on observations obtained with XMM-Newton, an ESA science mission with instruments and contributions directly funded by ESA Member States and NASA; and NuSTAR, a project led by the California Institute of Technology, managed by the Jet Propulsion Laboratory, and funded by the National Aeronautics and Space Administration. Data analysis was performed using the NuSTAR Data Analysis Software (NuSTARDAS), jointly developed by the ASI Science Data Center (SSDC, Italy) and the California Institute of Technology (USA).
\end{acknowledgements}



\bibliography{references}{}
\bibliographystyle{aasjournalv7}
%
%

\end{document}